\documentclass[twocolumn,prl,tightenlines,superscriptaddress]{revtex4-2}
\usepackage[breaklinks]{hyperref}
\usepackage{amsmath,amssymb}
\usepackage[utf8]{inputenc} 			
\usepackage[T1]{fontenc}				
\usepackage{graphicx}
\usepackage{float}
\usepackage[usenames,dvipsnames]{xcolor}
\usepackage{tikz}
\usepackage[normalem]{ulem}

\newcommand{\grad}{\nabla}
\renewcommand{\div}{\nabla\cdot}
\renewcommand{\vec}[1]{\mathbf{#1}}
\newcommand{\Gvec}[1]{\boldsymbol{#1}}

\newcommand*\colvec[3][]{
    \begin{pmatrix}\ifx\relax#1\relax\else#1\\\fi#2\\#3\end{pmatrix}
}

\newcommand{\nperp}{\vec n_\bot}
\newcommand{\dpar}{\partial_\parallel}
\newcommand{\dperp}{\partial_\bot}

\newcommand{\dd}{\mathrm{d}}
\newcommand{\bq}{\mathbf q}
\newcommand{\bk}{\mathbf k}
\newcommand{\bp}{\mathbf p}
\newcommand{\Om}{\Omega}
\newcommand{\sq}{\sqrt{D_xD_y}}
\newcommand{\dlx}{\delta\lambda_x}
\newcommand{\dM}{\delta M}
\newcommand{\tth}{\tilde\theta}

\begin{document}

\title{Emergent Symmetry in Constant-Density Flocks}

\author{Hugues Chat\'{e}}
\affiliation{Service de Physique de l'Etat Condens\'e, CEA, CNRS Universit\'e Paris-Saclay, CEA-Saclay, 91191 Gif-sur-Yvette, France}
\affiliation{Sorbonne Universit\'e, CNRS, Laboratoire de Physique Th\'eorique de la Mati\`ere Condens\'ee, 75005 Paris, France}
\author{Alexandre Solon}
\affiliation{Sorbonne Universit\'e, CNRS, Laboratoire de Physique Th\'eorique de la Mati\`ere Condens\'ee, 75005 Paris, France}

\date{\today}
\begin{abstract}
  In two dimensions, the ordered phase of constant-density flocks is
  described by a single angle field, which makes them arguably the
  simplest systems with a scale-free ordered phase of collective
  motion. We show that the equation governing the corresponding Nambu--Goldstone mode, 
  completed with terms previously discarded, possesses on large scales an
  invariance combining a shear and a Galilean boost. Broken at the bare level,
  this symmetry is restored at the unique attractive fixed point,
  and the associated Ward identities fix the
  scaling exponents exactly, as confirmed numerically.
\end{abstract}

\maketitle

A large part of the enthusiasm for active matter can be traced to the
fact that activity, a small-scale dissipation of energy, leads to heretofore unknown
non-equilibrium collective phenomena.
Examples now abound, 
but the first spectacular case is found in the seminal papers by Toner and
Tu~\cite{toner_long-range_1995,toner_flocks_1998}. Inspired by the
minimal model introduced by Vicsek and
collaborators~\cite{vicsek_novel_1995}, in which self-propelled
particles that locally align their velocities can move collectively,
they argued that such polar flocks exhibit true long-range order in
two dimensions, something forbidden by the Mermin--Wagner--Hohenberg
theorem in passive equilibrium
systems~\cite{mermin_absence_1966,hohenberg_existence_1967}.

Some of the fundamental properties of active systems remain poorly understood,
even in the case of polar flocks. After more than thirty years of effort, and 
despite recent propositions~\cite{jentsch_new_2024,ikeda_minimum_2024,chate_dynamic_2024}
and precise measurements~\cite{mahault_quantitative_2019}, the fundamental nature of their scale-free 
properties is still
actively debated~\cite{chen_inconvenient_2025,chate_comment_2025,chen_response_2025,chate_continuing_2025}.

Part of the difficulty to establish the scaling properties of polar
flocks is rooted in the fact that they necessarily involve both a
density and an order field.  A simpler class of systems showing
scale-free collective motion is that of constant-density flocks
(CDF). These were first considered by Toner in the form of
``Malthusian flocks''~\cite{toner_birth_2012} in which particles
divide and die on the fly so that the density is fixed by the carrying
capacity of the medium. On large scales, the density is effectively
constant and the system can be described at continuum level by an
equation governing a single vector field $\vec v$ that reads
\begin{align}
  \partial_t\vec v&+\lambda_1\left(\vec v\cdot\nabla\right) \vec v+\lambda_2(\div \vec v)\vec v+\lambda_3\nabla\left(|\vec v|^2\right)=\nonumber\\
                  &D_0\nabla^2\vec v +D_1\nabla\left(\div \vec v\right)+\left(a-b|\vec v|^2\right)\vec v+\sqrt{2T}\Gvec\xi  
                    \label{eq:full-v}
\end{align}
with $\Gvec\xi$ a Gaussian white noise with unit variance,
uncorrelated between different components. This constant-density
Toner--Tu equation is also the continuum description of non-reciprocal
XY
models~\cite{loos_long-range_2023,rouzaire_nonreciprocal_2025,popli_ordering_2025,dopierala_inescapable_2025},
in which classical XY spins interact, e.g., preferentially with the
spins in the direction where they point, up to lattice effects that
enter on large
scales~\cite{popli_ordering_2025,dopierala_inescapable_2025}.

Besse et al.~\cite{besse_metastability_2022} showed numerically that
CDF, governed by Eq.~(\ref{eq:full-v}), possess a scale-free
long-range ordered phase at low noise but that it is always metastable
to the nucleation of topological defects, the steady state then being
a disordered active foam of aster domains, also observed in lattice
non-reciprocal XY
models~\cite{popli_ordering_2025,dopierala_inescapable_2025}.

Although metastable, the scale-free long-range ordered phase of CDF
can be observed on long time and length scales, so its study remains
legitimate. In $d=2$, the hydrodynamic equation governing the
corresponding Nambu--Goldstone mode features only a scalar field
parameterizing the direction of order, and is thus particularly
simple. Nonetheless, there is still no consensus on the values of
$\chi$, $\zeta$, and $z$, the three scaling exponents involved, which
govern respectively the roughness of the field, the anisotropy of
fluctuations, and dynamical scaling.  Toner~\cite{toner_birth_2012}
first obtained the exact values $\chi=-1/5$, $\zeta=3/5$ and $z=6/5$
from scaling relations.  We later realized that a relevant term was
missing from the theory (the $\lambda_x$ term in
Eq.~(\ref{eq:theta-full})) and, based on a conjectured ``non-linear
Galilean invariance'', we obtained a different set of exponents
$\chi=-1/4$, $\zeta=3/4$ and
$z=5/4$~\cite{chate_dynamic_2024}. Numerically, we first concluded
that the fluctuations in the ordered phase of Eq.~(\ref{eq:full-v})
are compatible with the Toner scaling~\cite{besse_metastability_2022}
but later found a clear departure in larger-scale simulations and a good
agreement with the latter prediction~\cite{chate_dynamic_2024}. On the
theoretical side, Chen et al.~\cite{chen_inconvenient_2025} agreed
with the missing term but not with the revised exponent values,
arguing that it is impossible to find enough scaling relations to
solve the problem.  Recently, Jentsch and
Erzberger~\cite{jentsch_true_2026}, using the functional
non-perturbative renormalization
group~\cite{dupuis_nonperturbative_2021}, computed the exponents and
found their values equal to those we obtained. They nevertheless
argued that theirs is an approximate result that would break down for
higher-order refinements of their scheme. All this calls for
clarification.

In this Letter, we show that the ordered phase of CDF possesses an
emergent symmetry that combines a shear transformation and a Galilean
boost. The Ward identities associated with the symmetry provide the
scaling relation conjectured in~\cite{chate_dynamic_2024} and thus put
the scaling exponent values found there, $\chi=-1/4$, $\zeta=3/4$ and
$z=5/4$, on a firm footing.  Interestingly, the sheared Galilean
invariance is generically broken at bare level but restored at the
renormalization group (RG) fixed point. We show this by computing the
RG flow at one-loop order, where we find that the unique attractive
fixed point possesses the symmetry.  We confirm this numerically by
comparing how early (in length scales) simulations of the theory with
bare parameters belonging or not to the symmetric manifold reach the
scaling regime.

In the absence of a density field, the fluctuations in the ordered phase
are controlled by the Goldstone mode associated with the breaking of
rotational symmetry. In $d=2$, it is an angle field $\theta(\vec r,t)$
which, as argued in Ref.~\cite{chate_dynamic_2024}, evolves generically at second order
in gradients as
\begin{equation}
  \partial_t \theta+\lambda \dpar\theta = \div\left[D_\parallel\dpar\theta \vec n +D_\bot\dperp\theta \vec \nperp \right] +\sqrt{2\Delta}\xi
  \label{eq:theta-rot}
\end{equation}
with $\vec n=(\cos\theta,\sin\theta)$,
$\vec \nperp=(-\sin\theta,\cos\theta)$,
$\dpar\equiv \vec n\cdot \grad$, $\dperp\equiv \nperp\cdot \grad$ and
$\xi(\vec r,t)$ a unit-variance Gaussian white noise.

The trigonometric functions in $\vec n$ and $\vec \nperp$ contain
nonlinearities of all orders in $\theta$ while, on large scales, the
fluctuations of $\theta$ are expected to be small and thus governed by
the leading-order nonlinearities. Taking the global order to be in the
$x$-direction corresponding to $\theta=0$, we expand
Eq.~(\ref{eq:theta-rot}) in powers of $\theta$ to get
\begin{align}
  \label{eq:theta-full}
  \partial_t \theta+\lambda_y\partial_y(\theta^2)& +\lambda_x\partial_x(\theta^3)   = \left(D_x\partial_x^2+D_y\partial_y^2\right)\theta \nonumber \\  
  & -K\partial_x\partial_y(\theta^2)+M\partial_x^2(\theta^3) +\sqrt{2\Delta}\xi
\end{align}
with $\lambda_y=\lambda/2$, $\lambda_x=-\lambda/6$, $D_x=D_\parallel$,
$D_y=D_\bot$, $K=D_\bot-D_\parallel$ and $M=(D_\bot-D_\parallel)/3$. A
drift term $\lambda\partial_x\theta$ on the l.h.s. of
Eq.~(\ref{eq:theta-full}) has been removed by shifting to the comoving
frame $\vec r\to \vec r-\lambda t \vec {e_x}$. Note that, compared
to~\cite{chate_dynamic_2024}, we retain the $K$ and $M$ terms which,
as we will see in the following, are important. By contrast, we
drop a term $\partial_y^2(\theta^3)$ which is of the same order in
$\theta$ and derivatives as the $M$ term but will have a smaller
dimension at the fixed point.

{\it Sheared Galilean invariance.} Truncating Eq.~(\ref{eq:theta-rot})
into Eq.~(\ref{eq:theta-full}) breaks the rotational symmetry of the
original problem. Nevertheless, we find that this original symmetry
survives in the form of the following transformation
\begin{subequations}
\begin{align}
  \theta'(x,y,t) &= \theta(X,Y,t)+c \\
  X &= x+\kappa\left(c y-\lambda_yc^2t\right)\\ 
  Y &= y-2\lambda_yc t 
\end{align}
\label{eq:transfo}
\end{subequations}
which mixes a shear deformation and a Galilean boost in both the $x$
and $y$ directions with the same transformation parameter $c$ and
$\kappa\equiv K/D_y$. For $\kappa=1$, Eq.~(\ref{eq:transfo}) is a
rotation of amplitude $c$ with the term $-cx$ in $Y$ dropped and the
boost implements the rotation of the comoving frame. Using the chain
rule $\partial_x\theta'=\partial_X\theta$ and
$\partial_y\theta'=\partial_Y\theta+\kappa c \partial_X\theta$, one
can check that the combination
$\partial_y^2\theta-\kappa\partial_x\partial_y(\theta^2)+\tfrac{1}{3}\kappa^2\partial_x^2(\theta^3)$
is invariant under Eq.~(\ref{eq:transfo}) and ultimately that
Eq.~(\ref{eq:theta-full}) is invariant under the transformation if and
only if
\begin{equation}
  \label{eq:condition}
  3\lambda_x=-\kappa\lambda_y \quad {\rm and} \quad  3M=\kappa K \;.
\end{equation}
These conditions define the symmetric manifold. They are not satisfied
at bare level since the coefficients derived from
Eq.~(\ref{eq:theta-rot}) give $3\lambda_x=-\lambda_y$ and $3M=K$,
which both require $\kappa=1$ through Eq.~(\ref{eq:condition}),
whereas $\kappa=K/D_y=1-D_\parallel/D_\bot$. This breaking of the
symmetry is a consequence of the terms discarded in the truncation
leading to Eq.~(\ref{eq:theta-full}). Nevertheless, as we will show,
the symmetry is restored on large scales.

Let us first derive the consequences of the symmetry
Eq.~(\ref{eq:transfo}) for the renormalized coefficients, which we denote with a subscript $R$. 
We use a massive scheme in fixed dimension
$d=2$~\cite{callan_jr_broken_1970,symanzik_small_1970,parisi_field-theoretic_1980},
adding a mass term $-m\theta$ to Eq.~(\ref{eq:theta-full}) that serves
as an infrared regulator, cutting fluctuations beyond the scale
$\ell=\sqrt{D_{y,R}/m}$. The renormalized parameters can be read
from the inverse response function $\Gamma_c(\vec q,\omega)$ of the
fluctuations $\varphi$ around a uniformly tilted state
$\theta=c+\varphi$~\cite{medina_burgers_1989,frey_two-loop_1994,canet_nonperturbative_2011}
which reads
\begin{align}
\Gamma_c(\vec q,\omega)= & -i\omega+D_{x,R}q_x^2+D_{y,R}q_y^2+m \nonumber \\
&+ c\big(2i\lambda_{y,R}\,q_y-2K_R\,q_xq_y\big) \nonumber \\
& +c^2\big(3i\lambda_{x,R}\,q_x+3M_R\,q_x^2\big)+\dots
\label{eq:Gammac}
\end{align}
where the dots denote terms of higher order in $q_{x,y}$ and/or
$c$. The RG time is $s=\log\ell$ and the anomalous dimension of a
parameter $P$ is $\eta_P=\dd\log P_R/\dd s$ at fixed bare
parameters. As discussed
before~\cite{toner_birth_2012,chate_dynamic_2024}, $\eta_\Delta=0$
exactly to all orders because all nonlinearities are derivatives
so noise-vertex corrections are $O(q^2)$~\cite{tauber_critical_2014}.

On the symmetric manifold, the sheared Galilean invariance imposes
additional relations on the anomalous dimensions. As detailed in End
Matter, the Ward identity derived from Eq.~(\ref{eq:transfo}) for the
inverse response function reads
\begin{equation}
  \label{eq:ward-response}
  \Gamma_c(\vec q,\omega)=\Gamma_0(q_x,q_y-\kappa c q_x,\omega-2\lambda_yc q_y+\kappa\lambda_y c^2 q_x)
\end{equation}
and imposes that 
\begin{equation}
  \eta_{\lambda_y}=\eta_{\lambda_x}=0,\qquad \eta_K=\eta_M=\eta_{D_y}.
  \label{eq:ward-dimensions}
\end{equation}
so that only $D_x$ and $D_y$ renormalize independently.

At a non-trivial fixed point with $\kappa\neq 0$, invariant under a
rescaling $y\to by$, $x\to b^{\zeta} x$, $t\to b^{z}t$,
$\theta\to b^{\chi}\theta$, Eq.~(\ref{eq:ward-dimensions}) gives the
scaling relations $\chi+z-1=0$ and $2\chi+z-\zeta=0$ which, together
with the relation $z-2\chi-1-\zeta=0$ coming from $\eta_\Delta=0$,
yielding the exponent values found in \cite{chate_dynamic_2024}: $\chi=-1/4$, $\zeta=3/4$, and $z=5/4$.

{\it RG flow.} We have seen so far that a fixed point obeying the symmetry
necessarily has the above exponent values, but it remains to be shown that such a
fixed point exists and is attractive. We do this at one-loop order in
End Matter, obtaining the flow
\begin{align}
\frac{\dd D_y}{\dd s}&\!=\!D_y\big(z\! - \!2\! + \!\eta_{D_y}\big), &
\frac{\dd D_x}{\dd s}&\!=\!D_x\big(z\! - \!2\zeta\! + \!\eta_{D_x}\big),\nonumber\\
\frac{\dd \lambda_y}{\dd s}&\!=\!\lambda_y\big(\chi\! + \!z\! - \!1\! + \!\eta_{\lambda_y}\big), &
\frac{\dd \lambda_x}{\dd s}&\!=\!\lambda_x\big(2\chi\! + \!z\! - \!\zeta\! + \!\eta_{\lambda_x}\big),\nonumber\\
\frac{\dd K}{\dd s}&\!=\!K\big(\chi\! + \!z\! - \!\zeta\! - \!1\! + \!\eta_K\big), \!\!\!\!\!\!&
\frac{\dd M}{\dd s}&\!=\!M\big(2\chi\! + \!z\! - \!2\zeta\! + \!\eta_M\big),\nonumber\\
\frac{\dd \Delta}{\dd s}&\!=\!\Delta\big(z\! - \!\zeta\! - \!1\! - \!2\chi\big), & \frac{\dd m}{\dd s}&\!=\! m\big(z \!-\! 2\!+\!\eta_{D_y}\big),
\label{eq:dimflow}
\end{align}
where all the anomalous dimensions, whose expressions are given in
Eq.~(\ref{eq:etas}), are proportional to $\Delta$.

Including the mass, Eq.~(\ref{eq:theta-full}) has eight parameters and
four dimensions (for $x$, $y$, $t$ and $\theta$), so the RG flow can
 be conveniently rewritten in terms of four dimensionless
parameters, which we choose proportional to the nonlinearities
\begin{equation}
  \begin{aligned}
    &\hat\lambda_y=\frac{\lambda_y\Delta^{1/2}}{(4\pi)^{1/2}D_y^{3/4}D_x^{1/4}m^{1/2}},
    &&\hat\lambda_x=\frac{\lambda_x\Delta}{4\pi D_xD_y^{1/2}m^{1/2}},\\
    &\hat{K}=\frac{K\Delta^{1/2}}{(4\pi)^{1/2}(D_xD_y)^{3/4}},
    &&\hat{M}=\frac{M\Delta}{4\pi D_y^{1/2}D_x^{3/2}}.
    \label{eq:couplings}
  \end{aligned}
\end{equation}
The flow~\eqref{eq:dimflow} can then be rewritten as
\begin{equation}
  \begin{aligned}
    \frac{\dd \hat\lambda_y}{\dd s}&=\hat\lambda_y\Big(1-\tfrac{15}{4}\hat\lambda_y^2+\hat{K}^2-\tfrac92 \hat{M}\Big),\\
    \frac{\dd \hat\lambda_x}{\dd s}&=\hat\lambda_x\Big(1-3\hat\lambda_y^2+3\hat{K}^2-15\hat{M}\Big),\\
    \frac{\dd \hat{K}}{\dd s}&=\hat{K}^3-\tfrac94 \hat\lambda_y^2\hat{K}-\tfrac{15}{2}\hat{K}\hat{M}-9\hat\lambda_y\hat\lambda_x,\\
    \frac{\dd \hat{M}}{\dd s}&=9\hat\lambda_x^2-\tfrac32 \hat\lambda_y^2\hat{M}+3\hat{K}^2\hat{M}-18\hat{M}^2 .
  \end{aligned}
  \label{eq:flow}
\end{equation}
This makes clear that, if starting with $\hat{K}=\hat{M}=0$ as
in~\cite{chate_dynamic_2024,chen_inconvenient_2025}, $K$ and $M$ are
switched on immediately, with
$\dd \hat{K}/\dd s=-9\hat\lambda_y\hat\lambda_x$ and
$\dd \hat{M}/\dd s=9\hat\lambda_x^2$ and thus need to be included in
the theory. We can also note that the flow is symmetric under
$(\hat \lambda_y,\hat K)\to -(\hat \lambda_y,\hat K)$ and
$(\hat \lambda_y,\hat \lambda_x)\to -(\hat \lambda_y,\hat \lambda_x)$
reflecting the inversion symmetries of Eq.~(\ref{eq:theta-full}).

\begin{figure}
  \centering
  \includegraphics[width=\columnwidth]{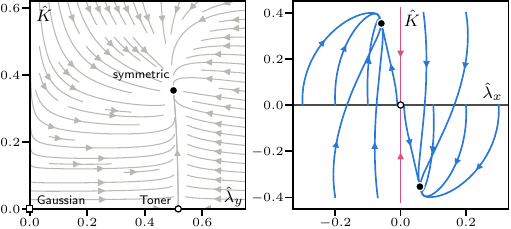}
  \caption{Left: Streamlines of the RG flow Eq.~(\ref{eq:flow}) on the
    symmetric manifold parameterized by $\hat\lambda_y$ and $\hat K$
    with $\delta_1=\delta_2=0$. Right: Projection of the flow on
    $\hat\lambda_x$ and $\hat K$ for several initial conditions with
    $\hat\lambda_y^0=2/\sqrt{15}$ and $\hat M^0=0$. The magenta
    trajectory with $\hat\lambda_x^0=0$ converges to the Toner fixed
    point. All other initial conditions (blue lines), including with
    $\hat K^0=0$, converge to the symmetric fixed point.}
  \label{fig:flow}
\end{figure}

The fixed points of the flow Eq.~(\ref{eq:flow}) are easily
enumerated. One finds five unstable fixed points: the Gaussian one
where all couplings vanish, the Toner fixed
point~\cite{toner_birth_2012} where only $\hat\lambda_y$ is non-zero,
and fixed points where $\hat\lambda_y=\hat K=0$,
$\hat\lambda_x=\hat K=0$ and $\hat\lambda_x=\hat M=0$. Up to the sign
symmetries, there is only one stable fixed point
$(\hat\lambda_y,\hat\lambda_x,\hat{K},\hat{M})=\big(\tfrac12,\,-\tfrac{\sqrt2}{24},\,\tfrac{\sqrt2}{4},\,\tfrac1{24}\big)$
that lies on the symmetric manifold. We have thus shown that the
sheared Galilean invariance is generically an emergent symmetry of
constant-density flocks which fixes the scaling exponents to the
values $\chi=-1/4$, $\zeta=3/4$ and $z=5/4$ conjectured
in~\cite{chate_dynamic_2024}.

{\it Approach to the fixed point.} To distinguish the convergence
toward the symmetric manifold from the convergence to the fixed point
within the manifold, it is convenient to introduce 
\begin{equation}
  \label{eq:distance}
  \delta_1\equiv \hat\lambda_x+\tfrac13 \hat\lambda_y\hat{K},\qquad
  \delta_2\equiv \hat{M}-\tfrac13 \hat{K}^2,
\end{equation}
which are dimensionless and vanish on the symmetric manifold. Using
Eq.~(\ref{eq:flow}) they flow as
\begin{equation}
\begin{aligned}
\frac{\dd\delta_1}{\dd s}&=\delta_1\big[1-6\hat\lambda_y^2-2\hat{K}^2-15\delta_2\big]+\hat\lambda_y\hat{K}\,\delta_2,\\
\frac{\dd\delta_2}{\dd s}&=-\tfrac12\big(3\hat\lambda_y^2+8\hat{K}^2\big)\delta_2+9\big(\delta_1^2-2\delta_2^2\big).
\end{aligned}
\label{eq:devflow}
\end{equation}
The symmetric manifold $\delta_1=\delta_2=0$ is invariant, as expected
since the symmetry is preserved by the RG scheme.

Rewriting the full flow Eq.~(\ref{eq:flow}) in terms of
$(\delta_1,\delta_2,\hat\lambda_y,\hat K)$, the linearization at the
fixed point is block triangular and yields two stable directions
  with eigenvalues $-3/4$ and $-7/8$ associated with $\delta_1$ and
$\delta_2$ respectively and  two others with eigenvalues
$-(9\pm\sqrt{33})/8$ associated with the relaxation within the
symmetric manifold, which is thus slower, with the largest eigenvalue
giving the first correction-to-scaling exponent
$\omega=(9-\sqrt{33})/8\approx 0.41$. Note that these eigenvalues, as
well as the location of the fixed point, are not exact and depend on
the number of loops. By contrast, the scaling exponents are fixed by
the Ward identity irrespective of the RG scheme. Fig.~\ref{fig:flow}
shows the flow streamlines on the symmetric manifold as well as
projected trajectories with initial conditions out of the manifold.

\begin{figure}
  \centering
    \includegraphics[width=\columnwidth]{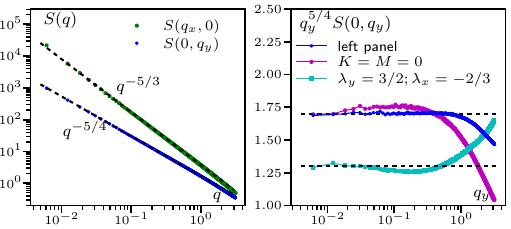}
  \caption{Structure factor $S(\bq)=\langle |\theta(\bq)|^2\rangle$
    measured in a semi-spectral integration of
    Eq.~(\ref{eq:theta-full}). Left: Both directions of $\bq$ compared
    with the computed exponents $1+\zeta+2\chi=z=5/4$ for $q_y$ and
    $(1+\zeta+2\chi)/\zeta=z/\zeta=5/3$ for $q_x$. Parameters:
    $\lambda_y=1$, $\lambda_x=-1/3$, $\Delta=0.1$, $K=D_y$, $M=D_y/3$,
    system size $L=1024$, time step $dt=0.1$ and lattice spacing $dx=1$.
    $D_x=0.065$ and $D_y=0.195$ chosen to optimize the convergence to
    the asymptotic regime. Right: Rescaled data for $\vec q=(0,q_y)$
    comparing different bare parameters: the ones of left panel (dark
    blue); $K=M=0$ with $D_x=0.1$, $D_y=0.25$ and other parameters
    unchanged (magenta); $\lambda_y=3/2$ and $\lambda_x=-2/3$ with
    other parameters as in the left panel (light blue).}
  \label{fig:numerics}
\end{figure}

In Fig.~\ref{fig:numerics}, we compare the behavior at large scales of
the structure factor $S(\bq)=\langle |\theta(\bq)|^2\rangle$, measured
in a semi-spectral integration of Eq.~(\ref{eq:theta-full}), with the
predicted scaling. Choosing parameters on the symmetric manifold and
$D_x$ and $D_y$ in order to minimize corrections to scaling, we find
an excellent agreement with the predicted scaling
(Fig.~\ref{fig:numerics}, left). Using other bare parameters, we see
that corrections to scaling are larger but the scaling exponents are
unchanged (Fig.~\ref{fig:numerics}(right)). This is rather direct
evidence of the emergence of the sheared Galilean symmetry on large
scales.

To summarize, we have shown that the hydrodynamic equation describing
constant-density flocks, complete with terms that were previously
discarded, possesses a sheared Galilean invariance when the conditions
of Eq.~(\ref{eq:condition}) are satisfied by the parameters. This
symmetry is the consequence of the rotational invariance of the
original system expressed in the comoving frame. This is a situation
akin to that of the KPZ equation~\cite{kardar_dynamic_1986} in which
the tilt symmetry is a remnant of rotational symmetry for a growing
interface~\cite{marsili_stochastic_1996}.

Here, the symmetry is broken at bare level but we demonstrated that it
is restored on large scales, at the RG fixed point. The scaling
exponents are fixed exactly by the symmetry at the values conjectured
in~\cite{chate_dynamic_2024} and also obtained in a functional RG
calculation~\cite{jentsch_true_2026}.

Let us mention that a transition was recently predicted between the
long-range ordered state described here and a quasi-long-range ordered
phase~\cite{grosvenor_renormalization_2026,jentsch_true_2026,sezik_large_2026}
where activity is irrelevant. This transition does not appear in our
RG treatment because it is seen only in the ``full''
Eq.~(\ref{eq:theta-rot}) and not in the truncated
Eq.~(\ref{eq:theta-full}). As discussed in~\cite{jentsch_true_2026},
it is in any case expected to happen in a parameter regime where the
ordered phase is already destroyed, even on short scales, by
topological defects~\cite{besse_metastability_2022} which are captured
neither by Eq.~(\ref{eq:theta-rot}) nor by Eq.~(\ref{eq:theta-full}).

Our findings raise the question of whether the sheared Galilean
invariance could be extended to other active systems and higher
dimensions or whether it is a mathematical specificity of the
system studied here.

\acknowledgements We thank L. Canet, B. Delamotte, M. Tissier and
N. Wschebor for stimulating discussions and a careful reading of the
manuscript.

\bibliography{refs.bib}

\begin{thebibliography}{32}%
\makeatletter
\providecommand \@ifxundefined [1]{%
 \@ifx{#1\undefined}
}%
\providecommand \@ifnum [1]{%
 \ifnum #1\expandafter \@firstoftwo
 \else \expandafter \@secondoftwo
 \fi
}%
\providecommand \@ifx [1]{%
 \ifx #1\expandafter \@firstoftwo
 \else \expandafter \@secondoftwo
 \fi
}%
\providecommand \natexlab [1]{#1}%
\providecommand \enquote  [1]{``#1''}%
\providecommand \bibnamefont  [1]{#1}%
\providecommand \bibfnamefont [1]{#1}%
\providecommand \citenamefont [1]{#1}%
\providecommand \href@noop [0]{\@secondoftwo}%
\providecommand \href [0]{\begingroup \@sanitize@url \@href}%
\providecommand \@href[1]{\@@startlink{#1}\@@href}%
\providecommand \@@href[1]{\endgroup#1\@@endlink}%
\providecommand \@sanitize@url [0]{\catcode `\\12\catcode `\$12\catcode
  `\&12\catcode `\#12\catcode `\^12\catcode `\_12\catcode `\%12\relax}%
\providecommand \@@startlink[1]{}%
\providecommand \@@endlink[0]{}%
\providecommand \url  [0]{\begingroup\@sanitize@url \@url }%
\providecommand \@url [1]{\endgroup\@href {#1}{\urlprefix }}%
\providecommand \urlprefix  [0]{URL }%
\providecommand \Eprint [0]{\href }%
\providecommand \doibase [0]{https://doi.org/}%
\providecommand \selectlanguage [0]{\@gobble}%
\providecommand \bibinfo  [0]{\@secondoftwo}%
\providecommand \bibfield  [0]{\@secondoftwo}%
\providecommand \translation [1]{[#1]}%
\providecommand \BibitemOpen [0]{}%
\providecommand \bibitemStop [0]{}%
\providecommand \bibitemNoStop [0]{.\EOS\space}%
\providecommand \EOS [0]{\spacefactor3000\relax}%
\providecommand \BibitemShut  [1]{\csname bibitem#1\endcsname}%
\let\auto@bib@innerbib\@empty
\bibitem [{\citenamefont {Toner}\ and\ \citenamefont
  {Tu}(1995)}]{toner_long-range_1995}%
  \BibitemOpen
  \bibfield  {author} {\bibinfo {author} {\bibfnamefont {J.}~\bibnamefont
  {Toner}}\ and\ \bibinfo {author} {\bibfnamefont {Y.}~\bibnamefont {Tu}},\
  }\bibfield  {title} {\bibinfo {title} {Long-range order in a two-dimensional
  dynamical {XY} model: how birds fly together},\ }\href@noop {} {\bibfield
  {journal} {\bibinfo  {journal} {Phys. Rev. Lett.}\ }\textbf {\bibinfo
  {volume} {75}},\ \bibinfo {pages} {4326} (\bibinfo {year}
  {1995})}\BibitemShut {NoStop}%
\bibitem [{\citenamefont {Toner}\ and\ \citenamefont
  {Tu}(1998)}]{toner_flocks_1998}%
  \BibitemOpen
  \bibfield  {author} {\bibinfo {author} {\bibfnamefont {J.}~\bibnamefont
  {Toner}}\ and\ \bibinfo {author} {\bibfnamefont {Y.}~\bibnamefont {Tu}},\
  }\bibfield  {title} {\bibinfo {title} {Flocks, herds, and schools: {A}
  quantitative theory of flocking},\ }\href@noop {} {\bibfield  {journal}
  {\bibinfo  {journal} {Physical review E}\ }\textbf {\bibinfo {volume} {58}},\
  \bibinfo {pages} {4828} (\bibinfo {year} {1998})}\BibitemShut {NoStop}%
\bibitem [{\citenamefont {Vicsek}\ \emph {et~al.}(1995)\citenamefont {Vicsek},
  \citenamefont {Czirók}, \citenamefont {Ben-Jacob}, \citenamefont {Cohen},\
  and\ \citenamefont {Shochet}}]{vicsek_novel_1995}%
  \BibitemOpen
  \bibfield  {author} {\bibinfo {author} {\bibfnamefont {T.}~\bibnamefont
  {Vicsek}}, \bibinfo {author} {\bibfnamefont {A.}~\bibnamefont {Czirók}},
  \bibinfo {author} {\bibfnamefont {E.}~\bibnamefont {Ben-Jacob}}, \bibinfo
  {author} {\bibfnamefont {I.}~\bibnamefont {Cohen}},\ and\ \bibinfo {author}
  {\bibfnamefont {O.}~\bibnamefont {Shochet}},\ }\bibfield  {title} {\bibinfo
  {title} {Novel type of phase transition in a system of self-driven
  particles},\ }\href@noop {} {\bibfield  {journal} {\bibinfo  {journal} {Phys.
  Rev. Lett.}\ }\textbf {\bibinfo {volume} {75}},\ \bibinfo {pages} {1226}
  (\bibinfo {year} {1995})}\BibitemShut {NoStop}%
\bibitem [{\citenamefont {Mermin}\ and\ \citenamefont
  {Wagner}(1966)}]{mermin_absence_1966}%
  \BibitemOpen
  \bibfield  {author} {\bibinfo {author} {\bibfnamefont {N.~D.}\ \bibnamefont
  {Mermin}}\ and\ \bibinfo {author} {\bibfnamefont {H.}~\bibnamefont
  {Wagner}},\ }\bibfield  {title} {\bibinfo {title} {Absence of ferromagnetism
  or antiferromagnetism in one-or two-dimensional isotropic {Heisenberg}
  models},\ }\href@noop {} {\bibfield  {journal} {\bibinfo  {journal} {Physical
  Review Letters}\ }\textbf {\bibinfo {volume} {17}},\ \bibinfo {pages} {1133}
  (\bibinfo {year} {1966})}\BibitemShut {NoStop}%
\bibitem [{\citenamefont {Hohenberg}(1967)}]{hohenberg_existence_1967}%
  \BibitemOpen
  \bibfield  {author} {\bibinfo {author} {\bibfnamefont {P.}~\bibnamefont
  {Hohenberg}},\ }\bibfield  {title} {\bibinfo {title} {Existence of long-range
  order in one and two dimensions},\ }\href@noop {} {\bibfield  {journal}
  {\bibinfo  {journal} {Physical Review}\ }\textbf {\bibinfo {volume} {158}},\
  \bibinfo {pages} {383} (\bibinfo {year} {1967})}\BibitemShut {NoStop}%
\bibitem [{\citenamefont {Jentsch}\ and\ \citenamefont
  {Lee}(2024)}]{jentsch_new_2024}%
  \BibitemOpen
  \bibfield  {author} {\bibinfo {author} {\bibfnamefont {P.}~\bibnamefont
  {Jentsch}}\ and\ \bibinfo {author} {\bibfnamefont {C.~F.}\ \bibnamefont
  {Lee}},\ }\bibfield  {title} {\bibinfo {title} {New {Universality} {Class}
  {Describes} {Vicsek}’s {Flocking} {Phase} in {Physical} {Dimensions}},\
  }\href@noop {} {\bibfield  {journal} {\bibinfo  {journal} {Phys. Rev. Lett.}\
  }\textbf {\bibinfo {volume} {133}},\ \bibinfo {pages} {128301} (\bibinfo
  {year} {2024})}\BibitemShut {NoStop}%
\bibitem [{\citenamefont {Ikeda}(2024)}]{ikeda_minimum_2024}%
  \BibitemOpen
  \bibfield  {author} {\bibinfo {author} {\bibfnamefont {H.}~\bibnamefont
  {Ikeda}},\ }\bibfield  {title} {\bibinfo {title} {Minimum scaling model and
  exact exponents for the {Nambu}-{Goldstone} modes in the {Vicsek} {Model}},\
  }\href@noop {} {\bibfield  {journal} {\bibinfo  {journal} {Phys. Rev. Lett.}\
  }\textbf {\bibinfo {volume} {133}},\ \bibinfo {pages} {258301} (\bibinfo
  {year} {2024})}\BibitemShut {NoStop}%
\bibitem [{\citenamefont {Chaté}\ and\ \citenamefont
  {Solon}(2024)}]{chate_dynamic_2024}%
  \BibitemOpen
  \bibfield  {author} {\bibinfo {author} {\bibfnamefont {H.}~\bibnamefont
  {Chaté}}\ and\ \bibinfo {author} {\bibfnamefont {A.}~\bibnamefont {Solon}},\
  }\bibfield  {title} {\bibinfo {title} {Dynamic scaling of two-dimensional
  polar flocks},\ }\href@noop {} {\bibfield  {journal} {\bibinfo  {journal}
  {Phys. Rev. Lett.}\ }\textbf {\bibinfo {volume} {132}},\ \bibinfo {pages}
  {268302} (\bibinfo {year} {2024})}\BibitemShut {NoStop}%
\bibitem [{\citenamefont {Mahault}\ \emph {et~al.}(2019)\citenamefont
  {Mahault}, \citenamefont {Ginelli},\ and\ \citenamefont
  {Chaté}}]{mahault_quantitative_2019}%
  \BibitemOpen
  \bibfield  {author} {\bibinfo {author} {\bibfnamefont {B.}~\bibnamefont
  {Mahault}}, \bibinfo {author} {\bibfnamefont {F.}~\bibnamefont {Ginelli}},\
  and\ \bibinfo {author} {\bibfnamefont {H.}~\bibnamefont {Chaté}},\
  }\bibfield  {title} {\bibinfo {title} {Quantitative assessment of the {Toner}
  and {Tu} theory of polar flocks},\ }\href@noop {} {\bibfield  {journal}
  {\bibinfo  {journal} {Phys. Rev. Lett.}\ }\textbf {\bibinfo {volume} {123}},\
  \bibinfo {pages} {218001} (\bibinfo {year} {2019})}\BibitemShut {NoStop}%
\bibitem [{\citenamefont {Chen}\ \emph
  {et~al.}(2025{\natexlab{a}})\citenamefont {Chen}, \citenamefont {Jentsch},
  \citenamefont {Lee}, \citenamefont {Maitra}, \citenamefont {Ramaswamy},\ and\
  \citenamefont {Toner}}]{chen_inconvenient_2025}%
  \BibitemOpen
  \bibfield  {author} {\bibinfo {author} {\bibfnamefont {L.}~\bibnamefont
  {Chen}}, \bibinfo {author} {\bibfnamefont {P.}~\bibnamefont {Jentsch}},
  \bibinfo {author} {\bibfnamefont {C.~F.}\ \bibnamefont {Lee}}, \bibinfo
  {author} {\bibfnamefont {A.}~\bibnamefont {Maitra}}, \bibinfo {author}
  {\bibfnamefont {S.}~\bibnamefont {Ramaswamy}},\ and\ \bibinfo {author}
  {\bibfnamefont {J.}~\bibnamefont {Toner}},\ }\bibfield  {title} {\bibinfo
  {title} {The inconvenient truth about flocks},\ }\href@noop {} {\bibfield
  {journal} {\bibinfo  {journal} {arXiv preprint arXiv:2503.17064}\ } (\bibinfo
  {year} {2025}{\natexlab{a}})}\BibitemShut {NoStop}%
\bibitem [{\citenamefont {Chaté}\ and\ \citenamefont
  {Solon}(2025{\natexlab{a}})}]{chate_comment_2025}%
  \BibitemOpen
  \bibfield  {author} {\bibinfo {author} {\bibfnamefont {H.}~\bibnamefont
  {Chaté}}\ and\ \bibinfo {author} {\bibfnamefont {A.}~\bibnamefont {Solon}},\
  }\bibfield  {title} {\bibinfo {title} {Comment on "{The} inconvenient truth
  about flocks" by {Chen} et al},\ }\href@noop {} {\bibfield  {journal}
  {\bibinfo  {journal} {arXiv preprint arXiv:2504.13683}\ } (\bibinfo {year}
  {2025}{\natexlab{a}})}\BibitemShut {NoStop}%
\bibitem [{\citenamefont {Chen}\ \emph
  {et~al.}(2025{\natexlab{b}})\citenamefont {Chen}, \citenamefont {Jentsch},
  \citenamefont {Lee}, \citenamefont {Maitra}, \citenamefont {Ramaswamy},\ and\
  \citenamefont {Toner}}]{chen_response_2025}%
  \BibitemOpen
  \bibfield  {author} {\bibinfo {author} {\bibfnamefont {L.}~\bibnamefont
  {Chen}}, \bibinfo {author} {\bibfnamefont {P.}~\bibnamefont {Jentsch}},
  \bibinfo {author} {\bibfnamefont {C.~F.}\ \bibnamefont {Lee}}, \bibinfo
  {author} {\bibfnamefont {A.}~\bibnamefont {Maitra}}, \bibinfo {author}
  {\bibfnamefont {S.}~\bibnamefont {Ramaswamy}},\ and\ \bibinfo {author}
  {\bibfnamefont {J.}~\bibnamefont {Toner}},\ }\bibfield  {title} {\bibinfo
  {title} {Response to the comment on "{The} inconvenient truth about flocks"
  by {Chaté} and {Solon}},\ }\href@noop {} {\bibfield  {journal} {\bibinfo
  {journal} {arXiv preprint arXiv:2505.21602}\ } (\bibinfo {year}
  {2025}{\natexlab{b}})}\BibitemShut {NoStop}%
\bibitem [{\citenamefont {Chaté}\ and\ \citenamefont
  {Solon}(2025{\natexlab{b}})}]{chate_continuing_2025}%
  \BibitemOpen
  \bibfield  {author} {\bibinfo {author} {\bibfnamefont {H.}~\bibnamefont
  {Chaté}}\ and\ \bibinfo {author} {\bibfnamefont {A.}~\bibnamefont {Solon}},\
  }\bibfield  {title} {\bibinfo {title} {Continuing the discussion on "{The}
  inconvenient truth about flocks" by {Chen} et al},\ }\href@noop {} {\bibfield
   {journal} {\bibinfo  {journal} {arXiv preprint arXiv:2506.13437}\ }
  (\bibinfo {year} {2025}{\natexlab{b}})}\BibitemShut {NoStop}%
\bibitem [{\citenamefont {Toner}(2012)}]{toner_birth_2012}%
  \BibitemOpen
  \bibfield  {author} {\bibinfo {author} {\bibfnamefont {J.}~\bibnamefont
  {Toner}},\ }\bibfield  {title} {\bibinfo {title} {Birth, death, and flight:
  {A} theory of malthusian flocks},\ }\href@noop {} {\bibfield  {journal}
  {\bibinfo  {journal} {Phys. Rev. Lett.}\ }\textbf {\bibinfo {volume} {108}},\
  \bibinfo {pages} {088102} (\bibinfo {year} {2012})}\BibitemShut {NoStop}%
\bibitem [{\citenamefont {Loos}\ \emph {et~al.}(2023)\citenamefont {Loos},
  \citenamefont {Klapp},\ and\ \citenamefont
  {Martynec}}]{loos_long-range_2023}%
  \BibitemOpen
  \bibfield  {author} {\bibinfo {author} {\bibfnamefont {S.~A.}\ \bibnamefont
  {Loos}}, \bibinfo {author} {\bibfnamefont {S.~H.}\ \bibnamefont {Klapp}},\
  and\ \bibinfo {author} {\bibfnamefont {T.}~\bibnamefont {Martynec}},\
  }\bibfield  {title} {\bibinfo {title} {Long-range order and directional
  defect propagation in the nonreciprocal {XY} model with vision cone
  interactions},\ }\href@noop {} {\bibfield  {journal} {\bibinfo  {journal}
  {Phys. Rev. Lett.}\ }\textbf {\bibinfo {volume} {130}},\ \bibinfo {pages}
  {198301} (\bibinfo {year} {2023})}\BibitemShut {NoStop}%
\bibitem [{\citenamefont {Rouzaire}\ \emph {et~al.}(2025)\citenamefont
  {Rouzaire}, \citenamefont {Pearce}, \citenamefont {Pagonabarraga},\ and\
  \citenamefont {Levis}}]{rouzaire_nonreciprocal_2025}%
  \BibitemOpen
  \bibfield  {author} {\bibinfo {author} {\bibfnamefont {Y.}~\bibnamefont
  {Rouzaire}}, \bibinfo {author} {\bibfnamefont {D.~J.}\ \bibnamefont
  {Pearce}}, \bibinfo {author} {\bibfnamefont {I.}~\bibnamefont
  {Pagonabarraga}},\ and\ \bibinfo {author} {\bibfnamefont {D.}~\bibnamefont
  {Levis}},\ }\bibfield  {title} {\bibinfo {title} {Nonreciprocal interactions
  reshape topological defect annihilation},\ }\href@noop {} {\bibfield
  {journal} {\bibinfo  {journal} {Phys. Rev. Lett.}\ }\textbf {\bibinfo
  {volume} {134}},\ \bibinfo {pages} {167101} (\bibinfo {year}
  {2025})}\BibitemShut {NoStop}%
\bibitem [{\citenamefont {Popli}\ \emph {et~al.}(2025)\citenamefont {Popli},
  \citenamefont {Maitra},\ and\ \citenamefont
  {Ramaswamy}}]{popli_ordering_2025}%
  \BibitemOpen
  \bibfield  {author} {\bibinfo {author} {\bibfnamefont {P.}~\bibnamefont
  {Popli}}, \bibinfo {author} {\bibfnamefont {A.}~\bibnamefont {Maitra}},\ and\
  \bibinfo {author} {\bibfnamefont {S.}~\bibnamefont {Ramaswamy}},\ }\bibfield
  {title} {\bibinfo {title} {Ordering and defect cloaking in nonreciprocal
  lattice {XY} models},\ }\href@noop {} {\bibfield  {journal} {\bibinfo
  {journal} {Phys. Rev. Lett.}\ }\textbf {\bibinfo {volume} {135}},\ \bibinfo
  {pages} {088303} (\bibinfo {year} {2025})}\BibitemShut {NoStop}%
\bibitem [{\citenamefont {Dopierala}\ \emph {et~al.}(2025)\citenamefont
  {Dopierala}, \citenamefont {Chaté}, \citenamefont {Shi},\ and\ \citenamefont
  {Solon}}]{dopierala_inescapable_2025}%
  \BibitemOpen
  \bibfield  {author} {\bibinfo {author} {\bibfnamefont {D.}~\bibnamefont
  {Dopierala}}, \bibinfo {author} {\bibfnamefont {H.}~\bibnamefont {Chaté}},
  \bibinfo {author} {\bibfnamefont {X.-q.}\ \bibnamefont {Shi}},\ and\ \bibinfo
  {author} {\bibfnamefont {A.}~\bibnamefont {Solon}},\ }\bibfield  {title}
  {\bibinfo {title} {Inescapable anisotropy of nonreciprocal {XY} models},\
  }\href@noop {} {\bibfield  {journal} {\bibinfo  {journal} {Phys. Rev. Lett.}\
  }\textbf {\bibinfo {volume} {135}},\ \bibinfo {pages} {088302} (\bibinfo
  {year} {2025})}\BibitemShut {NoStop}%
\bibitem [{\citenamefont {Besse}\ \emph {et~al.}(2022)\citenamefont {Besse},
  \citenamefont {Chaté},\ and\ \citenamefont
  {Solon}}]{besse_metastability_2022}%
  \BibitemOpen
  \bibfield  {author} {\bibinfo {author} {\bibfnamefont {M.}~\bibnamefont
  {Besse}}, \bibinfo {author} {\bibfnamefont {H.}~\bibnamefont {Chaté}},\ and\
  \bibinfo {author} {\bibfnamefont {A.}~\bibnamefont {Solon}},\ }\bibfield
  {title} {\bibinfo {title} {Metastability of {Constant}-{Density} {Flocks}},\
  }\href@noop {} {\bibfield  {journal} {\bibinfo  {journal} {Phys. Rev. Lett.}\
  }\textbf {\bibinfo {volume} {129}},\ \bibinfo {pages} {268003} (\bibinfo
  {year} {2022})}\BibitemShut {NoStop}%
\bibitem [{\citenamefont {Jentsch}\ and\ \citenamefont
  {Erzberger}(2026)}]{jentsch_true_2026}%
  \BibitemOpen
  \bibfield  {author} {\bibinfo {author} {\bibfnamefont {P.}~\bibnamefont
  {Jentsch}}\ and\ \bibinfo {author} {\bibfnamefont {A.}~\bibnamefont
  {Erzberger}},\ }\bibfield  {title} {\bibinfo {title} {True and {Quasi}
  {Long}-{Range} {Order} in {Malthusian} {Flocks}},\ }\href@noop {} {\bibfield
  {journal} {\bibinfo  {journal} {arXiv preprint arXiv:2608.05851}\ } (\bibinfo
  {year} {2026})}\BibitemShut {NoStop}%
\bibitem [{\citenamefont {Dupuis}\ \emph {et~al.}(2021)\citenamefont {Dupuis},
  \citenamefont {Canet}, \citenamefont {Eichhorn}, \citenamefont {Metzner},
  \citenamefont {Pawlowski}, \citenamefont {Tissier},\ and\ \citenamefont
  {Wschebor}}]{dupuis_nonperturbative_2021}%
  \BibitemOpen
  \bibfield  {author} {\bibinfo {author} {\bibfnamefont {N.}~\bibnamefont
  {Dupuis}}, \bibinfo {author} {\bibfnamefont {L.}~\bibnamefont {Canet}},
  \bibinfo {author} {\bibfnamefont {A.}~\bibnamefont {Eichhorn}}, \bibinfo
  {author} {\bibfnamefont {W.}~\bibnamefont {Metzner}}, \bibinfo {author}
  {\bibfnamefont {J.~M.}\ \bibnamefont {Pawlowski}}, \bibinfo {author}
  {\bibfnamefont {M.}~\bibnamefont {Tissier}},\ and\ \bibinfo {author}
  {\bibfnamefont {N.}~\bibnamefont {Wschebor}},\ }\bibfield  {title} {\bibinfo
  {title} {The nonperturbative functional renormalization group and its
  applications},\ }\href@noop {} {\bibfield  {journal} {\bibinfo  {journal}
  {Physics Reports}\ }\textbf {\bibinfo {volume} {910}},\ \bibinfo {pages} {1}
  (\bibinfo {year} {2021})}\BibitemShut {NoStop}%
\bibitem [{\citenamefont {Callan~Jr}(1970)}]{callan_jr_broken_1970}%
  \BibitemOpen
  \bibfield  {author} {\bibinfo {author} {\bibfnamefont {C.~G.}\ \bibnamefont
  {Callan~Jr}},\ }\bibfield  {title} {\bibinfo {title} {Broken scale invariance
  in scalar field theory},\ }\href@noop {} {\bibfield  {journal} {\bibinfo
  {journal} {Phys. Rev. D}\ }\textbf {\bibinfo {volume} {2}},\ \bibinfo {pages}
  {1541} (\bibinfo {year} {1970})}\BibitemShut {NoStop}%
\bibitem [{\citenamefont {Symanzik}(1970)}]{symanzik_small_1970}%
  \BibitemOpen
  \bibfield  {author} {\bibinfo {author} {\bibfnamefont {K.}~\bibnamefont
  {Symanzik}},\ }\bibfield  {title} {\bibinfo {title} {Small distance behaviour
  in field theory and power counting},\ }\href@noop {} {\bibfield  {journal}
  {\bibinfo  {journal} {Communications in Mathematical Physics}\ }\textbf
  {\bibinfo {volume} {18}},\ \bibinfo {pages} {227} (\bibinfo {year}
  {1970})}\BibitemShut {NoStop}%
\bibitem [{\citenamefont {Parisi}(1980)}]{parisi_field-theoretic_1980}%
  \BibitemOpen
  \bibfield  {author} {\bibinfo {author} {\bibfnamefont {G.}~\bibnamefont
  {Parisi}},\ }\bibfield  {title} {\bibinfo {title} {Field-theoretic approach
  to second-order phase transitions in two- and three- dimensional systems},\
  }\href@noop {} {\bibfield  {journal} {\bibinfo  {journal} {Journal of
  Statistical Physics}\ }\textbf {\bibinfo {volume} {23}},\ \bibinfo {pages}
  {49} (\bibinfo {year} {1980})}\BibitemShut {NoStop}%
\bibitem [{\citenamefont {Medina}\ \emph {et~al.}(1989)\citenamefont {Medina},
  \citenamefont {Hwa}, \citenamefont {Kardar},\ and\ \citenamefont
  {Zhang}}]{medina_burgers_1989}%
  \BibitemOpen
  \bibfield  {author} {\bibinfo {author} {\bibfnamefont {E.}~\bibnamefont
  {Medina}}, \bibinfo {author} {\bibfnamefont {T.}~\bibnamefont {Hwa}},
  \bibinfo {author} {\bibfnamefont {M.}~\bibnamefont {Kardar}},\ and\ \bibinfo
  {author} {\bibfnamefont {Y.-C.}\ \bibnamefont {Zhang}},\ }\bibfield  {title}
  {\bibinfo {title} {Burgers equation with correlated noise:
  {Renormalization}-group analysis and applications to directed polymers and
  interface growth},\ }\href@noop {} {\bibfield  {journal} {\bibinfo  {journal}
  {Phys. Rev. A}\ }\textbf {\bibinfo {volume} {39}},\ \bibinfo {pages} {3053}
  (\bibinfo {year} {1989})}\BibitemShut {NoStop}%
\bibitem [{\citenamefont {Frey}\ and\ \citenamefont
  {Täuber}(1994)}]{frey_two-loop_1994}%
  \BibitemOpen
  \bibfield  {author} {\bibinfo {author} {\bibfnamefont {E.}~\bibnamefont
  {Frey}}\ and\ \bibinfo {author} {\bibfnamefont {U.~C.}\ \bibnamefont
  {Täuber}},\ }\bibfield  {title} {\bibinfo {title} {Two-loop
  renormalization-group analysis of the {Burgers}–{Kardar}-{Parisi}-{Zhang}
  equation},\ }\href@noop {} {\bibfield  {journal} {\bibinfo  {journal} {Phys.
  Rev. E}\ }\textbf {\bibinfo {volume} {50}},\ \bibinfo {pages} {1024}
  (\bibinfo {year} {1994})}\BibitemShut {NoStop}%
\bibitem [{\citenamefont {Canet}\ \emph {et~al.}(2011)\citenamefont {Canet},
  \citenamefont {Chaté}, \citenamefont {Delamotte},\ and\ \citenamefont
  {Wschebor}}]{canet_nonperturbative_2011}%
  \BibitemOpen
  \bibfield  {author} {\bibinfo {author} {\bibfnamefont {L.}~\bibnamefont
  {Canet}}, \bibinfo {author} {\bibfnamefont {H.}~\bibnamefont {Chaté}},
  \bibinfo {author} {\bibfnamefont {B.}~\bibnamefont {Delamotte}},\ and\
  \bibinfo {author} {\bibfnamefont {N.}~\bibnamefont {Wschebor}},\ }\bibfield
  {title} {\bibinfo {title} {Nonperturbative renormalization group for the
  {Kardar}-{Parisi}-{Zhang} equation: {General} framework and first
  applications},\ }\href@noop {} {\bibfield  {journal} {\bibinfo  {journal}
  {Phys. Rev. E}\ }\textbf {\bibinfo {volume} {84}},\ \bibinfo {pages} {061128}
  (\bibinfo {year} {2011})}\BibitemShut {NoStop}%
\bibitem [{\citenamefont {Kardar}\ \emph {et~al.}(1986)\citenamefont {Kardar},
  \citenamefont {Parisi},\ and\ \citenamefont {Zhang}}]{kardar_dynamic_1986}%
  \BibitemOpen
  \bibfield  {author} {\bibinfo {author} {\bibfnamefont {M.}~\bibnamefont
  {Kardar}}, \bibinfo {author} {\bibfnamefont {G.}~\bibnamefont {Parisi}},\
  and\ \bibinfo {author} {\bibfnamefont {Y.-C.}\ \bibnamefont {Zhang}},\
  }\bibfield  {title} {\bibinfo {title} {Dynamic scaling of growing
  interfaces},\ }\href@noop {} {\bibfield  {journal} {\bibinfo  {journal}
  {Physical Review Letters}\ }\textbf {\bibinfo {volume} {56}},\ \bibinfo
  {pages} {889} (\bibinfo {year} {1986})}\BibitemShut {NoStop}%
\bibitem [{\citenamefont {Marsili}\ \emph {et~al.}(1996)\citenamefont
  {Marsili}, \citenamefont {Maritan}, \citenamefont {Toigo},\ and\
  \citenamefont {Banavar}}]{marsili_stochastic_1996}%
  \BibitemOpen
  \bibfield  {author} {\bibinfo {author} {\bibfnamefont {M.}~\bibnamefont
  {Marsili}}, \bibinfo {author} {\bibfnamefont {A.}~\bibnamefont {Maritan}},
  \bibinfo {author} {\bibfnamefont {F.}~\bibnamefont {Toigo}},\ and\ \bibinfo
  {author} {\bibfnamefont {J.~R.}\ \bibnamefont {Banavar}},\ }\bibfield
  {title} {\bibinfo {title} {Stochastic growth equations and reparametrization
  invariance},\ }\href@noop {} {\bibfield  {journal} {\bibinfo  {journal}
  {Reviews of Modern Physics}\ }\textbf {\bibinfo {volume} {68}},\ \bibinfo
  {pages} {963} (\bibinfo {year} {1996})}\BibitemShut {NoStop}%
\bibitem [{\citenamefont {Grosvenor}\ and\ \citenamefont
  {Patil}(2026)}]{grosvenor_renormalization_2026}%
  \BibitemOpen
  \bibfield  {author} {\bibinfo {author} {\bibfnamefont {K.~T.}\ \bibnamefont
  {Grosvenor}}\ and\ \bibinfo {author} {\bibfnamefont {S.~P.}\ \bibnamefont
  {Patil}},\ }\bibfield  {title} {\bibinfo {title} {On the {Renormalization}
  {Group} {Flow} of {Active} {Flocks}},\ }\href@noop {} {\bibfield  {journal}
  {\bibinfo  {journal} {arXiv preprint arXiv:2606.20552}\ } (\bibinfo {year}
  {2026})}\BibitemShut {NoStop}%
\bibitem [{\citenamefont {Sezik}\ and\ \citenamefont
  {Pruessner}(2026)}]{sezik_large_2026}%
  \BibitemOpen
  \bibfield  {author} {\bibinfo {author} {\bibfnamefont {E.}~\bibnamefont
  {Sezik}}\ and\ \bibinfo {author} {\bibfnamefont {G.}~\bibnamefont
  {Pruessner}},\ }\bibfield  {title} {\bibinfo {title} {Large {Spin}-{Wave}
  {Fluctuations} {Suppress} {Activity} in {Malthusian} {Flocks}},\ }\href@noop
  {} {\bibfield  {journal} {\bibinfo  {journal} {arXiv preprint
  arXiv:2608.05805}\ } (\bibinfo {year} {2026})}\BibitemShut {NoStop}%
\bibitem [{\citenamefont {Täuber}(2014)}]{tauber_critical_2014}%
  \BibitemOpen
  \bibfield  {author} {\bibinfo {author} {\bibfnamefont {U.~C.}\ \bibnamefont
  {Täuber}},\ }\href@noop {} {\emph {\bibinfo {title} {Critical dynamics: a
  field theory approach to equilibrium and non-equilibrium scaling behavior}}}\
  (\bibinfo  {publisher} {Cambridge University Press},\ \bibinfo {year}
  {2014})\BibitemShut {NoStop}%
\end{thebibliography}%

\onecolumngrid

\vspace{12pt}
\noindent\hrulefill \hspace{24pt} {\bf End Matter} \hspace{24pt} \hrulefill
\vspace{12pt}

\twocolumngrid

{\it One-loop RG flow.}
Let us first write Eq.~(\ref{eq:theta-full}), with an added external
force $h$ and regulator $-m\theta$, in Fourier space,
$\bp=(\bq,\omega)$:
\begin{align}
(-i\omega+\Om_\bq)\,\theta(\bp) =& \sqrt{2\Delta}\,\xi(\bp) + h(\bp) \nonumber \\
& -V(\bq)\,[\theta\theta](\bp) -U(\bq)\,[\theta\theta\theta](\bp) \label{eq:eqFourier}
\end{align}
with
\begin{align}
\Om_\bq &= D_xq_x^2+D_yq_y^2+m \nonumber\\
V(\bq) &=i\lambda_yq_y-K\,q_xq_y \nonumber\\
U(\bq) &=i\lambda_xq_x+M\,q_x^2.\nonumber
\end{align}
Above, $[\theta\theta](\bp)=\int_{\bp'}\theta(\bp')\theta(\bp-\bp')$,
and similarly for three fields, with
$\int_{\bp}=(2\pi)^{-3}\int \dd\bq \dd\omega$. $\xi$ is a unit
Gaussian white noise and $h$ is used to define the response
$\langle \theta(\bp)\rangle=\Gamma^{-1}(\bp)h(\bp)$.

We expand around a uniformly tilted state, $\theta=c+\varphi$ with a
uniform force $h=mc$ compensating the constant $-mc$ produced by the
mass term. The fluctuation field then obeys
\begin{align}
\big[-i\omega & + i\,\vec v_c\cdot\bq+\Om_\bq(c)\big]\varphi(\bp) = \sqrt{2\Delta}\,\xi(\bp)\nonumber\\ 
&+h(\bp) - V_c(\bq)\,[\varphi\varphi](\bp)-U(\bq)\,[\varphi\varphi\varphi](\bp), \label{eq:phi}
\end{align}
with 
\begin{subequations}
\begin{align}
\vec v_c &= (3\lambda_xc^2,\,2\lambda_yc) \;,\\
\Om_\bq(c) &= \Om_\bq-2cK\,q_xq_y+3c^2M\,q_x^2 \;,\\
V_c(\bq) &= V(\bq)+3ic\lambda_x\,q_x+3cM\,q_x^2  \;.
\end{align}
\label{eq:phi-coeffs}
\end{subequations}
At tree level, the inverse response function at zero frequency,
$\Gamma_c(\bq)$, is then
\begin{equation}
\Gamma_c(\bq)=\Om_\bq+2c\big[i\lambda_yq_y \!-\! Kq_xq_y\big]+3c^2\big[i\lambda_xq_x \!+\! Mq_x^2\big].
\label{eq:tree}
\end{equation}
The renormalized coefficients are defined by the same expansion of the
exact $\Gamma_c$:
\begin{align}
\Gamma_c(\bq)=&m \!+\! D_{x,R}q_x^2+D_{y,R}q_y^2+2c\big[i\lambda_{y,R}q_y \!-\! K_Rq_xq_y\big] \nonumber\\
&+3c^2\big[i\lambda_{x,R}q_x \!+\! M_Rq_x^2\big]+\dots
\label{eq:defR}
\end{align}
with dots denoting higher order terms in the expansion. The drift
$\mathbf v_c$ is absorbed inside the loops by shifting the loop
frequency $\omega-\mathbf v_c\cdot\bq$. At one loop, the resulting
frequency dependence only affects $c^2q_y^2$, outside the truncation.

Iterating Eq.~(\ref{eq:phi}) once and averaging over noise, we obtain
after performing the frequency integral
\begin{equation}
\Gamma_c(\bq)=\text{tree}-4V_c(\bq)\,J_c(\bq)+3U(\bq)\,\langle\varphi^2\rangle_c,
\label{eq:oneloop}
\end{equation}
with $\langle\varphi^2\rangle_c=\int_\bk\frac{\Delta}{\Om_\bk(c)}$ and
\begin{equation}
J_c(\bq)=\int_\bk V_c(\bq-\bk)\,\frac{\Delta}{\Om_\bk(c)\big[\Om_\bk(c)+\Om_{\bq-\bk}(c)\big]}.
\label{eq:J}
\end{equation}
We have omitted in Eq.~(\ref{eq:oneloop}) the $c$-independent part of
the tadpole, $3i\lambda_x\langle\varphi^2\rangle_0q_x$, a uniform
drift along $x$ that is removed by the choice of the comoving
frame. Every correction carries the factor $V_c(\bq)$ or $U(\bq)$,
which vanishes at $\bq=0$. Hence $m$ and the coefficient of $-i\omega$
are not renormalized. The corrections to the noise vertex are
$O(q^2)$, so $\Delta$ is not renormalized either.

\begin{figure}[h]
  \includegraphics[width=\columnwidth]{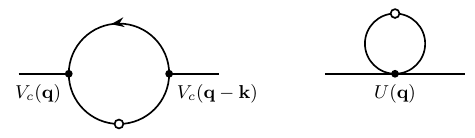}
  \caption{The two one-loop contributions to $\Gamma_c$ in
    Eq.~\eqref{eq:oneloop}. The line with arrow denotes the response
    function and the line with an open circle the correlation
    function.}
\label{fig:diag}
\end{figure}

Recall that $\kappa\equiv K/D_y$ and the deviations from the
symmetric manifold
\begin{equation}
\dlx\equiv\lambda_x+\frac{K\lambda_y}{3D_y},\qquad \dM\equiv M-\frac{K^2}{3D_y}.
\end{equation}
It is convenient to change variables
$\tilde q_y\equiv q_y-\kappa c\,q_x$. Then, Eq.~(\ref{eq:phi-coeffs})
reads
\begin{subequations}
\begin{align}
&\Om_\bq(c)=D_xq_x^2+D_y\tilde q_y^{\,2}+m+3c^2\dM\,q_x^2,\label{eq:shear}\\
&V_c(\bq)=i\lambda_y\tilde q_y-Kq_x\tilde q_y+3ic\,\dlx\,q_x+3c\,\dM\,q_x^2.
\end{align}
\end{subequations}
In these variables, the tilt enters $J_c$ and
$\langle\varphi^2\rangle_c$ only through $\dlx$ and $\dM$. On the
symmetric manifold, $\dlx=\dM=0$, we see directly that the tilted
loops are the untilted ones at wavevector $(q_x,q_y-\kappa cq_x)$,
as required by the Ward identity
Eq.~(\ref{eq:ward-response}).

To evaluate Eq.~(\ref{eq:oneloop}), we use 
\begin{equation}
\int_\bk\frac{1}{\Om_\bk}=\frac{\alpha}{4\pi\sq},\qquad
\int_\bk\frac{D_xk_x^2}{\Om_\bk^2}=\frac{\alpha}{8\pi\sq}
\label{eq:int}
\end{equation}
and other similar integrals, with $\alpha=\log(\Lambda^2/m)$,
$\Lambda$ being a UV cutoff. Note that $\alpha$ depends on the way the
theory is regularized in the UV but, in the RG flow, only
$\partial \alpha/\partial s=2-\eta_{D_y}$, independent of the regulator, enters.

The tadpole evaluates to
\begin{align}
&\langle\varphi^2\rangle_c=\langle\varphi^2\rangle_0-3c^2\dM\,\Delta\!\int_\bk\frac{k_x^2}{\Om_\bk^2}
=\langle\varphi^2\rangle_0+c^2\mu, \\
 & \langle\varphi^2\rangle_0=\frac{\Delta \alpha}{4\pi\sq},\qquad
  \mu\equiv-\frac{3 \alpha}{8\pi}\,\frac{\Delta\,\dM}{D_x^{3/2}D_y^{1/2}} .
\label{eq:tad}
\end{align} The loop diagrams give at zero wavevector
\begin{equation}
J_c(0)=\frac32\,c\,\dM\,\Delta\!\int_\bk\frac{k_x^2}{\Om_\bk^2}+O(c^3)=-\frac c2\,\mu .
\end{equation}
And at order $O(q)$ we get
\begin{align}
J_c(\bq)-J_c(0)&=i\frac{\rho}{4}\big(\lambda_y\tilde q_y+3c\,\dlx\,q_x\big) \nonumber \\
&=i\frac{\rho}{4}\big(\lambda_yq_y+3c\lambda_xq_x\big)
\label{eq:Jc}
\end{align}
with $\rho\equiv 3\Delta/(8\pi m\sq)$. We have dropped in
Eq.~(\ref{eq:Jc}) a term of order $c^2q_y$ that produces in $\Gamma_c$
a $c^2q_y^2$ term corresponding to a term $\partial_y^2(\theta^3)$
that is irrelevant at the fixed point.

Inserting $J_c$ and the tadpole into Eq.~\eqref{eq:oneloop}, we get an
expression for $\Gamma_c(\bq)$ from which we can identify the
renormalized coefficients
\begin{align}
&D_{y,R}=D_y+\rho\lambda_y^2, & D_{x,R}&=D_x+3M\langle\varphi^2\rangle_0, \nonumber\\
&\lambda_{y,R}=\lambda_y(1+\mu), & \lambda_{x,R}&=\lambda_x(1+3\mu), \label{eq:renorm}\\
&K_R=K(1+\mu)-3\rho\,\lambda_y\lambda_x, \!\!\!& M_R&=M(1+3\mu)+3\rho\,\lambda_x^2 . \nonumber
\end{align}

Differentiating Eq.~(\ref{eq:renorm}) with respect to $s$ then gives
\begin{align}
  &\eta_{D_y}=\frac{3}{4\pi}\frac{\lambda_y^2\Delta}{m\,D_y^{3/2}D_x^{1/2}},\nonumber\\
  & \eta_{D_x}=\frac{3}{2\pi}\frac{M\Delta}{D_x^{3/2}D_y^{1/2}},\nonumber\\
  &\eta_{\lambda_y}=\tfrac13\eta_{\lambda_x}=-\frac{3}{4\pi}\frac{\Delta\,\dM}{D_x^{3/2}D_y^{1/2}},  \label{eq:etas}
  \\
  &\eta_K=\eta_{\lambda_y}-\frac{9}{4\pi}\frac{\lambda_y\lambda_x\Delta}{K\,m\sq},\nonumber\\
  & \eta_M=3\eta_{\lambda_y}+\frac{9}{4\pi}\frac{\lambda_x^2\Delta}{M\,m\sq}.\nonumber
\end{align}
along with $\eta_\Delta=\eta_m=0$.  Finally, we rescale $y\to \ell y$,
$x\to \ell^\zeta x$, $t\to \ell^z t$ and $\theta\to \ell^\chi\theta$,
so that the coefficients are measured in units of the running scale, 
which gives the flow Eq.~(\ref{eq:dimflow}).

{\it Ward identities.}
We derive here the relations on anomalous dimensions
Eq.~(\ref{eq:ward-dimensions}) associated with the sheared Galilean
transformation Eq.~(\ref{eq:transfo}). The regularizing mass term is
not invariant under the transformation but, since it is linear, it
conveniently translates directly into the Ward identity for the
effective action
\begin{equation}
  \label{eq:ward-full}
  \Gamma[T_c\theta,T_c\tth]=\Gamma[\theta,\tth]+mc\!\int\!\tth
\end{equation}
for all $\theta,\tth,c,$ where $\tth$ is the response field and
$T_c$ is the operator realizing the transformation
Eq.~(\ref{eq:transfo})~\cite{tauber_critical_2014}.

The Ward identity for the inverse response function
Eq.~(\ref{eq:ward-response}) then follows by differentiating
Eq.~(\ref{eq:ward-full}) twice
\begin{equation}
\Gamma_c(\vec r, t)=\frac{\delta^2\Gamma[\theta,\tth]}{\delta\tth(\vec r,t)\,\delta\theta(\vec 0,0)}\bigg|_{\theta=c,\ \tth=0}.
\end{equation}
The non-renormalization of $\lambda_y$ follows from the Ward identity
Eq.~(\ref{eq:ward-response}) because, on the l.h.s. the renormalized
coefficients appear (as in Eq.~(\ref{eq:defR})) while on the r.h.s. the
bare $\lambda_y$ enters through the symmetry transformation. Looking
at the term that is linear in $q_y$ and $c$ then directly gives
$\lambda_{y,R}=\lambda_y$ and thus $\eta_{\lambda_y}=0$. Further,
comparing the terms in $c^2 q_x$, $c q_xq_y$ and $c^2 q_x^2$,
where $\kappa$ enters as a bare parameter, gives
$\lambda_{x,R}=-\kappa \lambda_y/3$, $K_R=\kappa D_{y,R}$ and
$M_R=\kappa^2 D_{y,R}/3$, giving all the equalities in
Eq.~(\ref{eq:ward-dimensions}).

\end{document}